# Cross-layer Design of CSMA/CA with Spectrum Sensing for Cognitive Radio Networks

Fotis Foukalas,
Dept. of Informatics and
Telecommunications,
National Kapodistrian University
of Athens
Athens, Greece
foukalas@di.uoa.gr

George T. Karetsos
Dept. of Information Technology
and Telecommunications,
Technology Education Institute
of Larissa
Larissa, Greece
karetsos@teilar.gr

Periklis Chatzimisios
Dept. of Information Technology
Technology Education Institute
of Thessaloniki
Thessaloniki, Greece
peris@it.teithe.gr

*Abstract*— We devise a cross-layer design (CLD) of carrier sensing multiple access with collision avoidance (CSMA/CA) at the medium access control (MAC) layer with spectrum sensing (SpSe) at the physical layer for cognitive radio networks (CRNs). The proposed CLD relies on a Markov chain model with a state pair containing both the SpSe and the CSMA/CA with exponential backoff from which we derive the transmission and collision probabilities. Due to the 2-dimensions of CSMA/CA model with exponential backoff, the resulted Markov chain is obtained with 3-dimensions. Simulation and numerical results are derived and illustrated highlighting the impact of SpSe in CSMA/CA with exponential backoff. The obtained results could be used as performance criteria to evaluate the performance of specific CRNs when they are deployed in a distributed coordination fashion that is prone to collisions.

## I. Introduction

In cognitive radio networks (CRNs), the channels' availability is manifested via spectrum sensing (SpSe) at the physical layer [1] and subsequently the packet transmission is accomplished through an appropriate medium access control (MAC) layer protocol [2]. It has been recognized that imperfect SpSe at the physical layer has an impact on the performance at the MAC layer as presented in [3]. To be specific, the impact of imperfect SpSe on the carrier sense multiple access with collision avoidance (CSMA/CA) protocol using a cross-layer performance analysis is highlighted by authors in [3]. On the other hand, authors in [4] propose a cross-layer design (CLD) between the SpSe mechanism at the physical layer and the MAC layer in general, i.e. without specifying a particular protocol, in which a constraint on collision probability dictates the operating characteristics of SpSe.

In this paper, we devise a cross-layer design (CLD) of SpSe with multi-channel SpSe at the physical layer with the CSMA/CA protocol considering exponential backoff. To this end, we model a 3-dimension Markov chain, which involves all parameters from both layers based on the concept proposed in [6], the Markov chain model of CSMA/CA presented in [7] and the Markov chain model of SpSe presented in [8]. Next, we derive the transmission and collision probabilities in saturation conditions for such a CRN. Notably, the proposed CLD for CRNs requires that the MAC layer of the CRN is able to support a distributed coordination method (DCF). For instance, such functionality is provided in IEEE 802.22 standard in two fashions:

- when dynamic frequency hopping (DFH) between adjacent base stations (BSs) is supported, where both SpSe and data transmission takes place from multiple adjacent cells and,
- when the point-to-point communication method (i.e. mesh CRN) is supported where the traffic flows among the customer premises equipments (CPEs) directly [5].

Our contribution compared with paper [3], it is summarized as follows: a) we provide a cross-layer design provision instead of cross-layer performance analysis and b) we additionally assume exponential backoff for the CSMA/CA. Finally, the proposed CLD model is evaluated using both numerical and simulation results.

This work is organized as follows. In section II the proposed cross layer design is introduced and the main parameters being involved are defined. Then in section III modeling of the augmented CSMA/CA protocol is taking place and through a bi-dimensional Markov chain the transmission and the collision probabilities are derived. Numerical results that illustrate the performance of the proposed scheme are provided in section IV and conclusions in section V.

## II. Cross-layer Design

We assume $n$ stations for the secondary network (SN) that could be either CPEs or BSs based on the type of the distributed coordination method that is implemented [5]. Each station is able to sense $c \in (1, C)$ channels of the primary network (PN) which in conjunction with the SN comprise the considered CRN. The spectrum sensing (SpSe) at the physical layer is taking place via an energy detection scheme over an additive white Gaussian noise (AWGN) channel. Energy detection is the most popular SpSe technique since it has low implementation complexity for unknown signals [1][9]. The energy detection scheme senses a signal to noise ratio (SNR) $\gamma$ of the primary signal with frequency $f_p$ for a sensing time $T_s$. We consider imperfect SpSe and thus four possible cases are introduced for the status of the PN perceived at the SN known as: detection, missed detection, false alarm and no false alarm with probabilities $P_d$, $(1-P_d)$, $P_f$ and $(1-P_f)$ [9]. The CSMA/CA protocol for packet transmission is considered at the MAC layer. In particular, the distributed coordination function (DCF) is considered in saturation conditions (i.e. each station has always a packet for transmission) with binary slotted exponential backoff of maximum backoff stage $m$ and minimum contention window of size $W$ [7]. We assume that for a slot at time $t$, a number of $C$ channels are being sensed and a node in the SN is able to transmit only in one channel per slot [8]. Thus, based on [7] and [8], we model both SpSe and CSMA/CA as discrete time Markov processes denoted as $s_t$ and $b_t$ respectively and in order to cross-layer design (CLD) the SpSe with CSMA/CA, we construct an augmented Markov chain with a state pair $(s_t, b_t)$ containing both the SpSe and the CSMA/CA states [6]. Fig.1 depicts the concept of the specific CLD where the backoff process $b_t$ at the MAC layer at time $t$ is changed based on the process of SpSe $s_t$ that takes place at the physical layer during the time slot $t$.

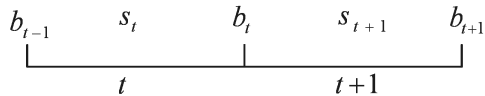

**Figure 1.** CSMA/CA and SpSe as discrete time Markov processes $b_t$ and $s_t$ respectively.

### III. MODELING

The CSMA/CA at the MAC layer is modeled as a bi-dimensional process $\{g(t), b(t)\}$ where $g(t)$ is the stochastic process of the backoff stage with size $i \in (0, m)$ and $b(t)$ is the stochastic process of the backoff timer with size $k \in (0, W_{i-1})$ where $W_i$ is equal to $2^i W$ [7].

The SpSe is modeled as a discrete time birth-death process $\{s(t)\}$ with size $s \in (0, C)$, where one channel per time slot $t$ is considered as active or idle with probability $a$ and $1-a$ respectively [8]. All channels are considered homogeneous i.e. with equal activity and the probability that a channel is active is $a(P_d) + (1-a)P_f$ due to imperfect SpSe. Thus, the CLD of SpSe with CSMA/CA that incorporates the channel's activity $a$, the spectrum sensing results $P_d$ and $P_f$, the backoff stage $m$ and the minimum contention window $W$ can be modeled as a three-dimensional stochastic process $\{g(t), b(t), s(t)\}$ with $i$, $k$ and $s$ dimensions respectively [6].

We denote the state transition probability as $P_{\{i,k,s\}\{j,l,u\}}$ that refers to the transition probability from state $\{g_t = i, b_t = k, s_{t-1} = s\}$ to state $\{g_{t+1} = j, b_{t+1} = l, s_t = u\}$ where $\{i,k,s\} \in [(0,m) \times (0, W_{i-1}) \times (0,C)]$, and $\{j,l,u\} \in [(0,m) \times (0, W_{i-1}) \times (0,C)]$. We illustrate in Fig.2 the derived Markov chain diagram at a backoff stage $i$ that shows the transition probabilities based on the SpSe results i.e. idle or busy derived within a discrete time slot $t$. For different backoff stages we follow the transitions defined in [7] that now include the results of the specific SpSe.

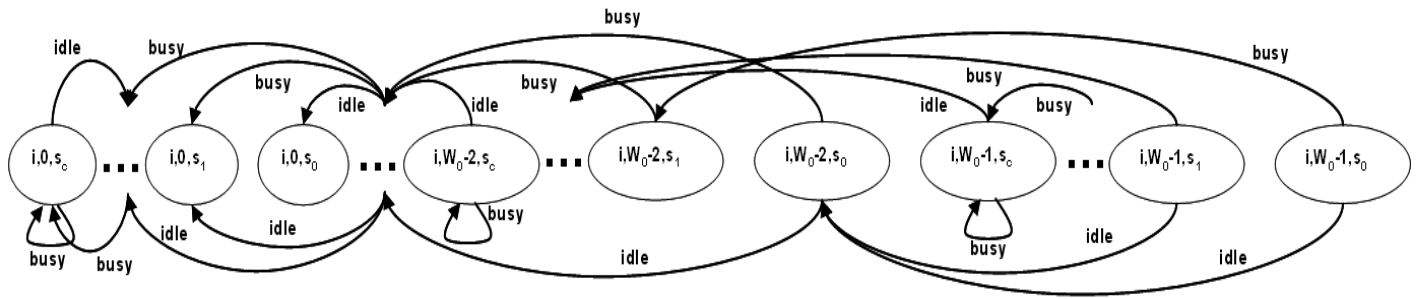

**Figure 2.** Markov chain diagram for a backoff stage $i$.

The state transition probability matrix is organized in a block form where matrix $A$ merges the stochastic process of the CSMA/CA protocol with an exponential backoff with $i \in (0,m)$ size given as follows:

$$A = \begin{bmatrix} \Pi_{\{0,0\}} & \cdots & \Pi_{\{0,W_0-1\}} \\ \cdots & \ddots & \cdots \\ \Pi_{\{m,0\}} & \cdots & \Pi_{\{m,W_m-1\}} \end{bmatrix} \quad (1)$$

which includes the sub-matrix $\Pi_{\{i,k\}\{j,l\}}$ that merges the stochastic process of SpSe $\{s(t)\}$ with size $s \in (0,C)$ which is given as follows

$$\Pi_{\{i,k\}\{j,l\}} = \begin{bmatrix} P_{\{i,k,s_0\}\{j,l,s_0\}} & \cdots & P_{\{i,k,s_0\}\{j,l,s_c\}} \\ \cdots & \ddots & \cdots \\ P_{\{i,k,s_c\}\{j,l,s_0\}} & \cdots & P_{\{i,k,s_c\}\{j,l,s_c\}} \end{bmatrix} . \quad (2)$$

This matrix represents a joint steady state probability that we denote here as $\pi_{\{i,k,s\}}$. It can be proved that this stationary distribution exists and is unique [6] and thus:

$$\pi_{\{i,k,s\}} := \lim_{t \to \infty} P(g_{t-1} = i, b_{t-1} = k, s_t = c) \quad . \quad (3)$$

The stationary distribution in (3) is a row vector $\pi$ defined as follows

$$\pi = [\pi_{\{0,0,0\}},...,\pi_{\{0,W_{i-1},0\}},...,\pi_{\{0,W_{i-1},C\}},...,$$
$$\pi_{\{m,0,0\}},...,\pi_{\{m,W_{i-1},0\}},...,\pi_{\{m,W_{i-1},C\}}] \quad (4)$$

for which the following holds

$$\sum_{i,k,s} \pi_{\{i,k,s\}} = \sum_{i=0}^{m} \sum_{k=0}^{W_{i-1}} \sum_{s=0}^{C} \pi_{\{i,k,s\}} = 1 \quad . \quad (5)$$

Based on this joint stationary distribution, we can derive the transmission and collision probabilities. In particular, a transmission occurs when the backoff timer is equal to zero i.e. $k = 0$ and when at least one channel is idle i.e. $s \in (0, C-1)$. Thus, the transmission probability is defined as follows:

$$\tau = \sum_{i=0}^{m} \sum_{s=0}^{C-1} \pi_{\{i,0,s\}} \quad . \quad (6)$$

Since the SpSe and the CSMA/CA are mutually independent, the transmission probability can be expressed as

$$\tau = \sum_{i=0}^{m} b_{i,0} \sum_{s=0}^{C-1} s_c \quad (7)$$

where $b_{i,0}$ and $s_c$ are the stationary probabilities of the CSMA/CA and the SpSe processes respectively [7][8]. The collision probability $p_c$ that is beget at the SN when at least one of the $n-1$ remaining stations of the PN transmits on the same channel is derived as follows [7, eq.(9)]:

$$p_c = 1 - (1-\tau)^{n-1} \quad . \quad (8)$$

## IV. NUMERICAL AND SIMULATION RESULTS

We have derived both numerical and simulation results in order to validate the proposed CLD. We assume that the sensed SNR is equal to $\gamma_p = -15dB$ and the sensing time equal to $T_s = 2ms$ for primary channels with frequency $f_p = 6MHz$ that is used in the case of the IEEE 802.22 in which TV bands are sensed [5]. Fig.3 shows the transmission probability $\tau$ versus the number of stations $n$ for different probabilities of detection $P_d$ and minimum contention window $W$ for a constant backoff stage $m = 3$. The results obtained considering one channel i.e. $C = 1$ with an activity probability equal to $a = 0.5$. The solid lines depict the case of $m = 3$ and $W = 32$ and the dashed lines depict the case of $m = 3$ and $W = 64$.

The simulation results are depicted without lines and they demonstrate that the analytical model is very accurate since its results almost coincide with the simulation ones. From the figure is obvious that a low probability of detection e.g. $P_d = 0.1$, results in high transmission probability $\tau$. Furthermore, high contention window values e.g. $W = 64$ (dashed lines) result in a lower transmission probability $\tau$, although a high contention window value gets the transmission probability lower. Fig.4 shows the collision probability $p_c$ considering the same parameterization as previously. Since the collision probability $p_c$ is proportional to the transmission probability $\tau$, this outcomes are inline with the previous ones.

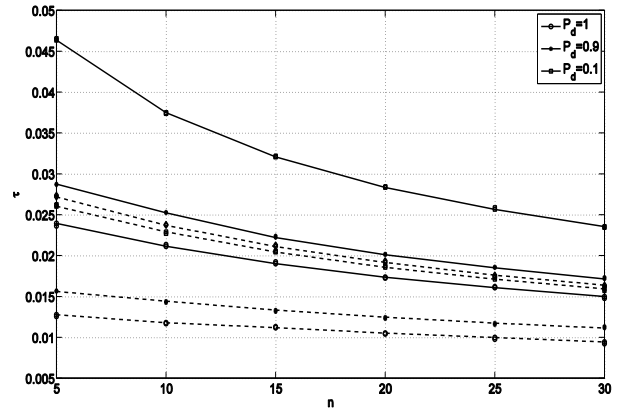

**Figure 3.** Transmission probability $\tau$ vs. number of stations $n$ for different probabilities of detection $P_d$ with backoff stage $m = 3$ and contention window $W = 32$ (solid lines) and with $m = 3$ and $W = 64$ (dashed lines).

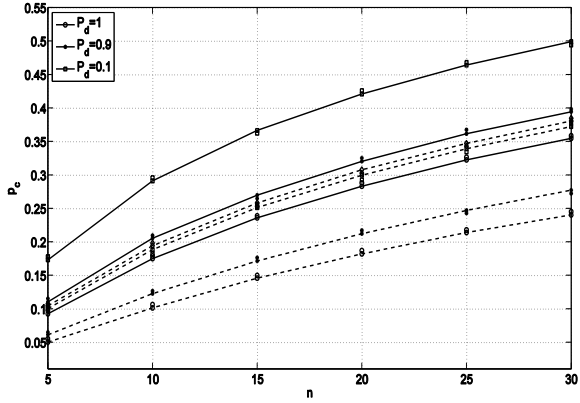

**Figure 4.** Collision probability $p_c$ vs. number of stations $n$ for different probabilities of detection $P_d$ with backoff stage $m=3$ and contention window $W=32$ (solid lines) and with $m=3$ and $W=64$ (dashed lines).

In Fig.5 and Fig.6, we show the transmission probability $\tau$ and the collision probability $p_c$ versus the number of stations $n$ for different probabilities of detection $P_d$ with a backoff stage equal to $m=5$ for a constant minimum contention window $W=32$. We also depict the simulation results without lines. It can be observed that the changes in the backoff stage induce less decrease in the transmission probability $\tau$ than the one induced by the changes in the contention window. A similar outcome is inferred from Fig.6 which shows the corresponding collision probability.

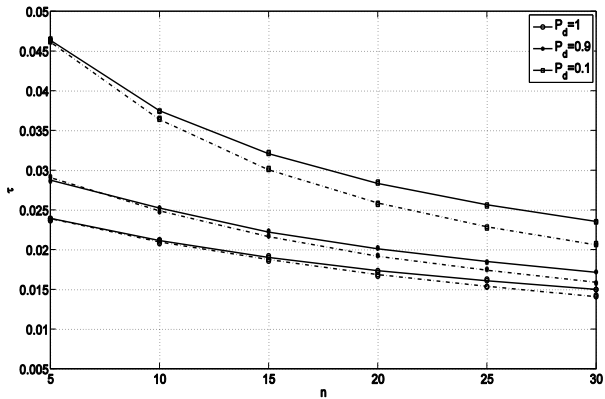

**Figure 5**. Transmission probability $\tau$ vs. number of stations $n$ for different probabilities of detection $P_d$ with backoff stage $m=3$ and contention window $W=32$ (solid lines) and with $m=5$ and $W=32$ (dotted dashed lines).

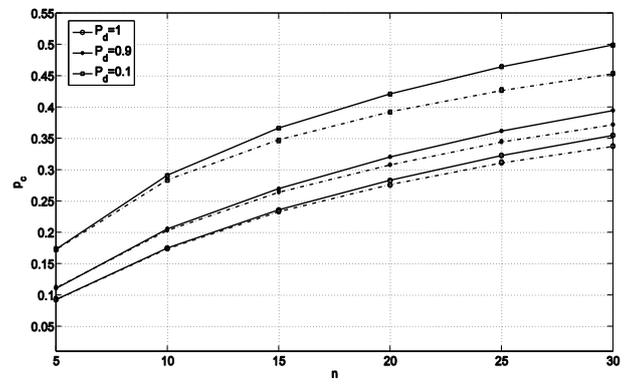

**Figure 6.** Collision probability $p_c$ vs. number of stations $n$ for different probabilities of detection $P_d$ with backoff stage $m=3$ and contention window $W=32$ (solid lines) and with $m=5$ and $W=32$ (dotted dashed lines).

Fig.7 shows the transmission probability $\tau$ versus the number of stations $n$ for different values of channels' activity $a$ and number of channels $C$. The results are obtained considering a probability of detection equal to $P_d=0.5$, a backoff stage equal to $m=3$ and a contention window equal to $W=32$. The solid lines depict the case of $a=0$, $a=0.5$, $a=0.8$ and $C=1$, the dashed lines depict the case of $C=3$ and the dotted dashed lines the case of $C=6$ considering the same activities for all cases. We also depict the simulation results without lines. Obviously, a high probability of activity e.g. $a=0.8$ results in a lower transmission probability $\tau$. Furthermore, for a high number of sensed channels e.g. $C=6$, the transmission probability $\tau$ increases.

Fig.8 shows the collision probability $p_c$ considering the same parameterization as previously and the outcomes are proportional to those derived for the transmission probability.

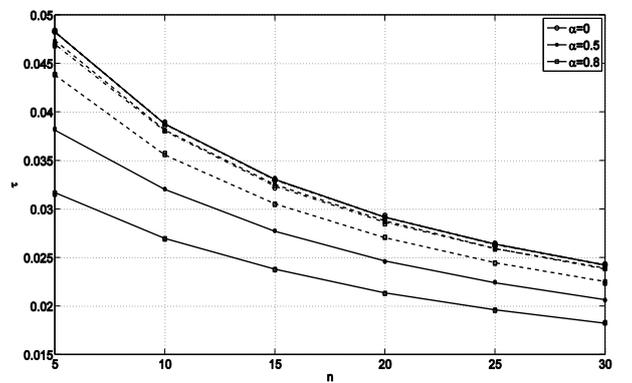

**Figure 8.** Transmission probability $\tau$ vs. number of stations $n$ for different activities $a$ with $C=1$ channels (solid lines), with $C=3$ channels (dashed lines) and with $C=6$ channels (dotted dashed lines).

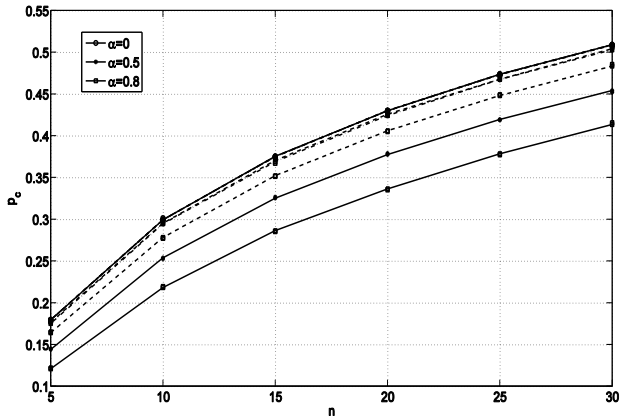

**Figure 9.** Collision probability $p_c$ vs. number of stations $n$ for different activities $a$ with $C=1$ channels (solid lines), with $C=3$ channels (dashed lines) and with $C=6$ channels (dotted dashed lines).

## V. CONCLUSIONS

We have devised a cross-layer design of spectrum sensing and CSMA/CA from which we have derived the transmission and collision probabilities. We rely on discrete time Markov chain model with a state pair for modeling both the spectrum sensing and the CSMA/CA protocol. This model results in a joint stationary probability which incorporates the parameters of spectrum sensing and CSMA/CA at the physical and the medium access control layer respectively. Thus the impact of SpSe in CSMA/CA protocol is realized which can be deployed in CRNs that support distributed coordination function such as the one found in the MAC of the IEEE protocol families.

ACKNOWLEDGMENT

This research has been co-financed by the European Union (European Social Fund - ESF) and Greek national funds through the Operational Program "Education and Lifelong Learning" of the National Strategic Reference Framework (NSRF) - Research Funding Program: ARCHIMEDES III. Investing in knowledge society through the European Social Fund.